# H-bonds in Crambin:
# Coherence in an α-helix

February 28, 2023

Stanley Nicholson##
David D. L. Minh[#]

Robert Eisenberg[*]
bob.eisenberg@gmail.com
Department of Applied Mathematics
Illinois Institute of Technology

[#]Department of Applied Mathematics
Illinois Institute of Technology

[##]Department of Chemistry
Illinois Institute of Technology

[*] Corresponding Author

# Abstract

We applied coherence analysis–used by engineers to identify linear interactions in stochastic systems–to molecular dynamics simulations of crambin, a thionin storage protein found in Abyssinian cabbage. A key advantage of coherence over other analyses is that it is robust, independent of the properties, or even the existence of probability distributions often relied on in statistical mechanics. For frequencies between 0.391 GHz and 5.08 GHz (corresponding reciprocally to times of 2.56 ns and 0.197 ns), the displacement of oxygen and nitrogen atoms across α-helix H-bonds are strongly correlated, with a coherence greater than 0.9; the secondary structure causes the H-bonds to effectively act as a spring. Similar coherence behavior is observed for covalent bonds and other noncovalent interactions including H-bonds in β-sheets and salt bridges. In contrast, arbitrary pairs of atoms that are physically distant have uncorrelated motions and negligible coherence. These results suggest that coherence may be used to objectively identify atomic interactions without subjective thresholds such as H-bond lengths angles and angles. Strong coherence is also observed between the average position of adjacent leaves (groups of atoms) in an α-helix, suggesting that the harmonic analysis of classical molecular dynamics can successfully describe the propagation of allosteric interactions through the structure.

# Introduction

Molecular dynamics (MD) simulations of proteins and other biological macromolecules describe the motions of their atoms in far greater detail than could be imagined by the founders of molecular biology and biophysics. Large improvements in the resolution and reliability of static structures have been combined with exponential increases in computing power to produce enormous amounts of simulation results. These data are a treasure trove for scientists, but are also, as treasures often are, a weighty burden. The exhaustive detail in MD trajectories easily overwhelms our ability to understand how the motion drives biological function.

The analysis of biomolecular MD trajectories typically combines visualization, data science, and traditional concepts from physical chemistry. Interpretation often starts with watching movies of cartoons that are reduced models of structure. General methods of data science such as clustering and dimensionality reduction are needed because the detail of atomic trajectories can hide the simplicity of reduced models. These methods are widely used to categorize configurations and define important motions seen in visualizations. Exploratory data analysis is often followed by more careful computation of thermodynamic properties such as expectation values and free energy differences between different conformations or kinetic properties of their transitions (see some newer literature[3, 4, 5, 6], but earlier references can be found). However, reduced models are difficult to identify and construct with certainty.

Fortunately, stochastic analysis[7] provides the coherence function, which allows the identification of a certain type of reduced model <u>without any assumptions</u> about the distribution or statistical properties of atomic motion. If the coherence function (defined below in Eq. (3)) is one, then the input and output of a stochastic process is related by a linear system, such as a spring. Use of the coherence function is well-established in engineering. Estimators of coherence are described in classic literature[7, 8] and are used with little change in modern literature[9-13]. One of us (Eisenberg) has used these methods in physiology to analyze the electrical structure of cells and tissues[14, 15], as has been reviewed[16, 17-19].

This work is far from the first to apply methods of stochastic analysis to MD trajectories. The time series of atomic positions in MD simulations are nearly as wild as the white noise signals used in engineering; trajectories follow wildly irregular paths of thermal motion that can be described as chaotic or stochastic or both[20], with the caveat that the motion is conditioned by the structures in which they occur. Dynamical properties of stochastic processes such as autocorrelation functions have been extensively investigated[4, 5] and are available in programs including TRAVIS[6, 21, 22]. However, to our knowledge, joint coherent thermal motions of atoms in proteins have not previously been studied in this engineering tradition.

Although we did not have high expectations that it would work, we decided that a good starting point would be to evaluate the coherence between pairs of atoms. Our expectations were tempered by several factors. First, all atoms in a protein interact with many other atoms, not only a single neighbor. Thus, predicting the motion of any single atom (the output) could require the positions of multiple atoms (the inputs). Second, conformational changes in proteins occur on all time scales from femtoseconds to seconds, which could disrupt coherence at any time scale. Finally, protein conformational changes show no visual signs of linearity, at least as we looked at them. Hence, we only really expected to find high coherence between pairs of atoms in covalent bonds. Nonetheless, we decided that it was straightforward enough to calculate pairwise coherence for other types of interactions.

We decided to start with a specific case important in the history of molecular biology: the H–bond that links nitrogen and oxygen within an α-helix of a protein, as discovered and defined by Linus Pauling[23]. As an initial system, we selected crambin, a small protein used for thionin storage by Abyssinian cabbage. Crambin has been a favored object of study by structural and molecular biologists for nearly 50 years as described in an extensive literature sampled in [24, 25, 26, 27]. We chose crambin because we knew it to be rigid and expected it to have particularly well-defined H-bonds. We thought that would be rigid because it is a small protein with three disulfide bonds[24] and has an especially high-resolution structure in X-ray and neutron diffraction experiments (Protein Data Bank identifier 1CRN[27]).

We wanted to select a rigid protein because we thought that atom pairs in rigid structures are more likely to have high coherence than in flexible proteins. Solids, an extreme case of rigidity, may have coherence over long distances. The best examples are in semiconductors, in which the entire quantum mechanical band structure arises from macroscopic periodicity[28] and of course in metal conductors, in which quantum states are delocalized over the entire length of a wire, even many kilometers. Moreover, starting from Einstein's 1907 theory for the specific heat of crystalline solids, which arguably gave birth to solid-state physics [29], there is a long history of treating solids as a lattice of harmonic oscillators. Thus, we thought that a rigid protein, crambin, would be one of the proteins most likely to exhibit some coherence.

To our delight, we found that that H-bonds in the main α-helix of crambin are nearly a linear system! The nitrogen and oxygen atoms across the H-bond share the same power—i.e., their power spectra are indistinguishable—and are well-correlated, with a coherence function greater than 0.9 from 0.391 GHz to 5.08 GHz (corresponding reciprocally to real time averages over 2.56 ns and 0.197 ns). Moreover, the corresponding frequency function shows no sign of damping and appears to be that of a rigid body (i.e., rigid bond). We repeat that this conclusion makes no assumptions concerning the existence let alone other properties of the distributions of atomic trajectories, as even nonparametric methods must assume[30]. Evidently, nonlinear forces and multiple interactions do not corrupt the linear system connecting the nitrogen and oxygen atoms of this H-bond and the other interactions we study here (salt bridges, covalent bonds, and the leaves – groups of atoms – of the main α-helix itself).

The rest of this paper follows the following plan. The Theory section is intended to introduce stochastic methods from linear systems analysis to a biomolecular simulation audience; a reader familiar with coherence analysis may skip the section without loss of continuity. The section starts with definitions of a linear system, frequency response function, and coherence function. The coherence function is contrasted with the Pearson Correlation Coefficient. We then describe statistical estimators that robustly estimate coherence. Finally, we describe analytical results for a driven damped spring. Following the Theory section, we include a Methods section with enough detail so that our simulations and estimates may be readily reproduced. Subsequently, Results of the calculations are presented along with controls that test the validity of this novel approach. We recognize that MD is usually performed with periodic boundary conditions that are generally not present in the real world, even in crystals—as discussed by many authors including p. 35-38 of [31]—and that digital and discrete stochastic analysis can be seriously corrupted by artifactual periodicities[7, 8, 32, 33, 34]. A wide variety of controls have been performed to ensure that our conclusions are robust. Next, we present a Discussion about the meaning of the linear systems that we have identified and how coherence analysis could be extended to other systems, even to sets of atoms in a 'Coherence Field Theory'. Finally, we summarize our Conclusions.

# Theory

**Linear systems.** The phrase *linear system* is defined in innumerable engineering textbooks. Following the MIT tradition, we cite Guillemin[5] and Bush[6] and use the term to describe a system with the following relationship between input and output:

(1) if two inputs are applied, output is the sum of the response to each individual input;

(2) if an input is multiplied by a constant *k*, then the output is multiplied by the same constant.

Fig. 1 shows a schematic of a linear system.

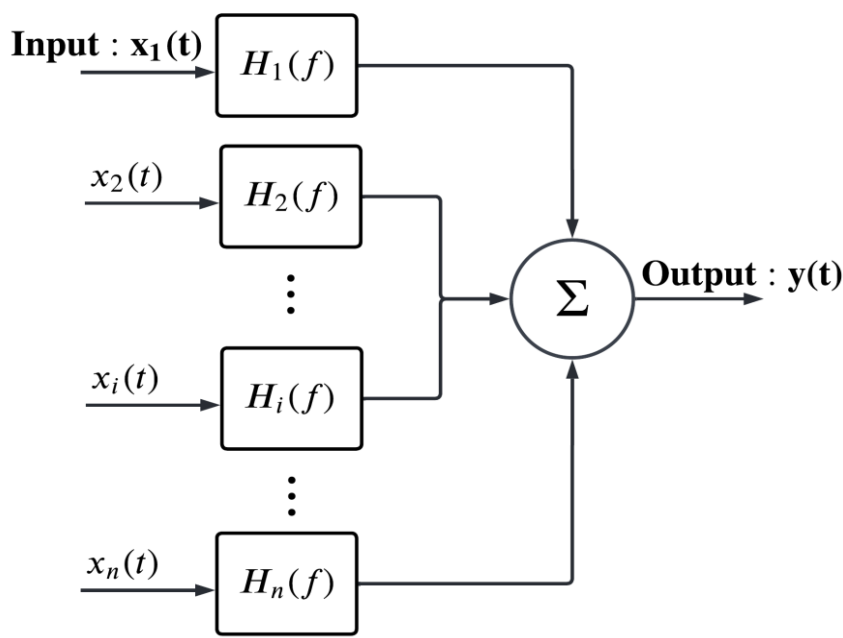

*Figure 1 Schematic of a linear system*. $x_i(t)$ are inputs, $H_i(f)$ are frequency response functions, and $y(t)$ is the output.

In this paper, we will focus on single input-output systems. A single input-output description is natural for a chemical bond, which is classically defined as a structure between a pair of atoms[35]. We do not deal with the mathematical[36] and chemical difficulties[37] in defining a two-body property like an ionic bond for a large three-dimensional system of charges, periodic or not, in which the system extends 'to infinity', with unavoidable consequences[18, 38]. We hope that a two-body analysis will be as successful in our chemical context as two-body (input-output) analysis has been of engineering devices. Chemical bonds are not isolated structures, particularly in biological contexts, but neither are engineering devices. Atoms in molecules are usually connected to more than one atom, but engineered devices also feature multiple connections. As in biological systems, some engineering devices are embedded in large three-dimensional systems. Some of these, such as computer memory, are even periodic in structure.

**The Frequency Response Function.** For a causal linear system with a single input $x(t)$, the output $y(t)$ is given by the convolution of the input with the *impulse response* $h(t)$. A causal system[39] is a system where the output depends on past and current inputs but not on future inputs—i.e., the output $y(t_0)$ depends only on the input $x(t)$ for values of $t \leq t_0$. The convolution is,

$$y(t) = \int_0^\infty h(\tau)x(t-\tau)d\tau. \tag{1}$$

This convolution is discussed at length in Ch. 2 of Bendat & Piersol[7].

A linear system converts a sinusoidal input to a sinusoidal output of the same frequency (after it settles down and transients disappear) but with an amplitude and phase that depend on frequency. The frequency response function $H(f)$ is defined based on the Fourier transform of Eq. (1). We refer here to a one-sided Fourier transform (i.e., Laplace transform), since signals start at some definite time. The frequency response function relates $X(f)$, the Fourier transform of the input $x(t)$, to the $Y(f)$, the Fourier transform of the output, $y(t)$, and is the Fourier transform of the impulse response $h(t)$:

$$H(f) = \frac{Y(f)}{X(f)} \tag{2}$$

The functions $X(f)$, $Y(f)$, and $H(f)$ have complex values with real and imaginary parts, or equivalently magnitude and phase.

While a sinusoid signal at one frequency can estimate the properties of the linear system at that frequency, measurements at all frequencies are needed to define the system. Fortunately, an input signal of white noise, a Brownian stochastic process[40, 41, 42], enables the simultaneous study of all frequencies. Measurements over a wide range of frequencies can be made much more quickly with wide-band stochastic signals. Speed is of particular importance in biological systems which often deteriorate in experimental situations [15, 17, 43, 44]. In MD simulations, evaluation of all frequencies at once reduces the compute time required to perform linear systems analysis.

Although, in principle, correlation functions in the time and frequency domains provide equivalent information, the frequency domain is more useful for identifying properties of linear systems[7, 8, 11, 13, 45]. Underlying mathematical issues of working in the time domain are discussed at length in [46]. The frequency domain is useful because it is based on the eigenfunctions of linear systems, namely sinusoids. Sinusoids are fundamentally special because they are natural solutions of the differential equations that describe causal networks. Chosen properly, sinusoids are orthogonal, with amplitudes and magnitudes measured at different frequencies independent of each other (in the asymptotic limit of long-lasting measurements).

Another benefit of following the engineering tradition of working in the frequency domain[13, 46] is the availability of the extensively validated and documented software including the Signal Processing Toolbox[47] in MATLAB[48]. These methods are discussed at length in Bendat and Piersol[7, 8] and Otnes and Enochson[49]. Other methods may improve on classical engineering approaches, but it is unrealistic to expect them to be available, validated, and documented for some time. In fact, most recent reviews use classical methods[7] in single input-output systems like ours[10, 13]. Relevant promising approaches in two (and three) dimensions are described in [8, 11, 12, 32, 45, 50].

While not too common in classical MD, there actually is a long history of using the frequency response function $H(f)$ in materials science and biochemistry. Measurements of $H(f)$ have been the basis of much of material science for more than a century[17, 51], as has been carefully explained[52]. Electrical ('impedance') measurements of the electrical properties of proteins were once used extensively[53] before protein structures were available but are rarely mentioned nowadays. Impedance measurements often report the spatial average of properties of the entire protein and so are not particularly helpful when the interesting properties arise in a tiny fraction of the protein, like an enzyme active or binding site, or ion channel.

Both the impulse response $h(t)$ and frequency response function $H(f)$ depend on the structure and parameters of the components of a particular system. The damped (Hookean) spring is the classical archetype of a linear system. For the damped spring, the mechanical parameters of $H(f)$ are the damping, mass, and most importantly the spring constant arranged in a series structure. However, many different structures have the same frequency response function $H(f)$[54]. Thus, the topology of the network and the values of its parameters cannot be determined without additional information. As has been reviewed[14, 15, 18, 44], structural (i.e., anatomical and histological) information has been used to determine circuit structures for many biological systems (skeletal muscle, cardiac muscle, epithelia, lens of the eye, syncytia in general, etc.[14, 15, 19, 44, 55]).

**The Coherence Function.** A system can be determined to be linear based on the coherence function[17–29]. The coherence function is defined in terms of power and cross power spectral densities, $G_{xx}(f)$ and $G_{xy}(f)$, respectively, as,

$$C_{xy}(f) = \frac{|G_{xy}(f)|^2}{G_{xx}(f)G_{yy}(f)}. \tag{3}$$

Suppose that the input function is truncated such that $x_T(t) = x(t)$ if $|t| < T/2$; and 0 otherwise. Its Fourier transform is $X_T(f)$. The power spectral density is given by the real function,

$$G_{xx}(f) = 2 \lim_{T \to \infty} \frac{1}{T} \langle X_T(f)^* X_T(f) \rangle, \tag{4}$$

where the angular brackets indicate an expectation of the stochastic process. $G_{yy}(f)$ is defined similarly. On the other hand, the cross power spectral density $G_{xy}(f)$ is a complex function defined by,

$$G_{xy}(f) = 2 \lim_{T \to \infty} \frac{1}{T} \langle X_T(f)^* Y_T(f) \rangle. \tag{5}$$

By relating the spectral densities to the Fourier transform of the autocorrelation function, it can be shown[7] that for a linear system, $G_{yy}(f) = |H(f)|^2 G_{xx}(f)$ and $G_{xy}(f) = H(f)G_{xx}(f)$ (see our eq. (9) & (10)). We have then,

## The Coherence Theorem

$$C_{xy}(f) = \frac{|H(f)G_{xx}|^2}{G_{xx}(f)|H(f)|^2 G_{xx}} = 1$$

when (6)

$$H(f) = \frac{Y(f)}{X(f)}$$

The coherence theorem shows that the coherence function $C_{xy}(f)$ gives the fraction of the output signal power $G_{yy}(f)$ that is linearly related to the input signal power $G_{xx}(f)$. If other signals contribute to the output, the coherence function is less than one. If and only if the coherence function is unity, the output is entirely determined by the input. Then, but only then, is the frequency function independent of anything else. ***The coherence function is one only if the output and input share power: all the power in the output comes from the power in the input.*** In more precise mathematical language, 'comes from' means 'is the result of a linear system $H(f)$ acting on $G_{xx}(f)$.' If the coherence function is much less than one, then the frequency function can be estimated, but is not meaningful. These results do not depend on the existence or properties of any distributions of the trajectories.

*These results depend only on the existence of the Fourier transforms involved.* The Fourier transforms (etc.) in fact can be defined and computed for a wide range of stochastic processes, many much more irregular than the continuous trajectories of Brownian stochastic processes. Brownian trajectories in actual applications have compact support (because the duration of recording is finite) and so convergence issues as $t \to \infty$ do not arise, in contrast to their importance in pure mathematics with its focus on unlimited domains. Jumps are permitted in trajectories, provided they are not too dense. They must not include too much area (i.e., power), to speak informally. These issues are discussed in standard textbooks of stochastic processes[40] as well as the powerful emerging methods[41, 42, 56, 57] exploiting the KH (Kurzweil-Henstock) integral to define stochastic differential equations and their integral transforms like the Fourier and Laplace transforms[42, 57].

**Coherence Function versus the Pearson Correlation Coefficient**. The coherence function has similarities to the Pearson Correlation Coefficient $r$, or Pearson **ρ**, as it is named when computed from a sample. The Pearson Correlation Coefficient is a common statistic widely used to measure the linear correlation between two variables. Indeed, the coherence can be thought of as the Pearson $r$ in the frequency domain[58]. For both statistics, an absolute value of one indicates a linear relationship between the variables. While it does not specifically require time series data, the Pearson **ρ** can be computed between an input signal and an output signal with a delay. In contrast to the coherence function, however, Pearson's **ρ** is not robust; it is sensitive to outliers and depends on assumptions about the underlying distribution of the data. Hence, for stochastic time series data, it is preferable to estimate the coherence function.

**Estimation of the Frequency Response and Coherence Function.** In general, the frequency response and coherence function are unknown and must be estimated from time series data. In this paper, we will use a hat accent ˆ to denote the estimator of a quantity, such that $\hat{A}(f)$ is an estimator of $A(f)$. Estimates of quantities from samples have different properties from the underlying distribution. For example, the variance of an estimate depends on the number of samples, but the variance of the distribution does not. The expectation value of an estimate of a variable may not the same as the value of that variable in the underlying distribution, in which case the estimate is biased. The variance and bias of the estimate interact, and one is often improved at the expense of the other[7, 34, 59].

Estimates of the frequency response $H(f)$ and coherence function $C_{xy}(f)$ are particularly sensitive to the details of the estimation procedure. The frequency response and coherence function should be estimated based on the ratio of *estimated* power and cross power spectral densities. Specifically, the frequency response function should be estimated by,

$$\widehat{H}(f) = \frac{\widehat{G}_{xy}(f)}{\widehat{G}_{xx}(f)}, \tag{7}$$

and the coherence function by,

$$\widehat{C}_{xy}(f) = \frac{|\widehat{G}_{xy}(f)|^2}{\widehat{G}_{xx}(f)\widehat{G}_{YY}(f)}. \tag{8}$$

While it may seem reasonable to estimate the frequency function using $Y(f)/X(f)$ instead of the ratio of powers shown in Eq. (7), such estimators are known to produce serious artifacts; they are known to be so sensitive to contaminating noise that even the roundoff error of modern floating-point arithmetic can produce significant problems; see Ch. 6, particularly section 6.14 of [7] and more recent references[11, 13]. These artifacts may be difficult to diagnose. For example, as observed in earlier work[14, 15, 44, 60], the variance of the incorrect estimates can be independent of contaminating noise while the mean of the incorrect estimates can depend on the contaminating noise. Hence, the estimates starting with $Y(f)/X(f)$ should not be used, and estimates of $\widehat{G}_{xx}(f)$ and $\widehat{G}_{yy}(f)$, and $\widehat{G}_{xy}(f)$, should be computed individually ***before*** executing the division in the definitions of the estimates $\widehat{C}_{xy}(f)$ and $\widehat{H}(f)$.

How should the power and cross power spectral densities be estimated? Consider a system with the Fourier transform of the input $X(f)$, output $Y(f)$, and transfer function $H(f)$. All might be complex functions with real and imaginary parts, or equivalently magnitude and phase. The power spectral density of the input $x(t)$ is estimated by the real function,

$$\widehat{G}_{xx}(f) = 2 \lim_{T \to \infty} \tfrac{1}{T} E\{\widehat{X}_k^*(f,T)\,\widehat{X}_k(f,T)\} = \\ 2 \lim_{T \to \infty} \tfrac{1}{T} E\left\{\left|\widehat{X}_k^*(f,T)\,\widehat{X}_k(f,T)\right|^2\right\}, \tag{9}$$

where * indicates the complex conjugate, $E$ is the expectation of any stochastic process including $\widehat{X}(f,T)$; $k$ indexes the components of a *set* of Discrete (digital) Fourier Transforms $\widehat{X}_k(f,T)$ of the time series $\widehat{x}(t)$. $k$ also serves as the index for the set of time series $\widehat{y}_k(t)$. $\widehat{G}_{yy}(f)$ is similarly defined. $\widehat{G}_{xx}(f)$ and $\widehat{G}_{yy}(f)$ are real functions with zero imaginary part and zero phase. The cross power $\widehat{G}_{xy}(f)$ is a complex function estimated by,

$$\widehat{G}_{xy}(f) = 2 \lim_{T \to \infty} \tfrac{1}{T} E\{\widehat{X}_k^*(f,T)\widehat{Y}_k(f,T)\}. \tag{10}$$

Note that,
1) The expectations in Eqs. (9) and (10) use **sums over <u>independent</u> records,** taking the sums individually, one by one. Each is determined *separately* with***out*** term by term 'cancellation';
2) The expectations are asymptotes determined as $T \to \infty$.

The DFT analysis (discrete or digital or finite Fourier Transform analysis) necessary for Eqs. (9) and (10) should also account for the fact that measurement data are often small, discrete, and truncated. Molecular dynamics trajectories are not an exception. While they may include a large amount of data, they are discretized because they use a finite time step and configurations are typically stored at regular periods. They are truncated because they are performed for a finite duration. The procedures defined in classical references to deal with these issues are rather scattered and hard to locate and integrate[7, 8, 32, 33, 34]; Section 6.12 – 6.13 and p. 128, 176, and p. 180 of [7]. Fortunately, they are implemented in software including the Signal Processing Toolbox[59] package of MATLAB[48]. To highlight the importance of using validated software and because DFT issues are not well-known to the molecular simulation community, we will describe these issues in more detail below. However, the reader uninterested in the underlying mathematical issues can move to the next section without loss of continuity.

In other applications, including most simulations, computers allow good approximation to continuous functions. Truncation and sampling errors can be unimportant because the density (and number) of discrete data points can be so very large. In other words, many types of errors can be 'diluted out' (as the chemists say) by the large numbers of samples. However, DFT is usually based on a small number of samples, on the order of 2048 in many cases, and is often as small as (the recommended default value of) 256 in the software we use. In that case, artifacts produced by the limited number of samples can be very serious, both discretization artifacts and truncation artifacts[61]. Choices for the number of DFT points depend on the desired frequency resolution: $[number\ of\ DFT\ points]$ equals $[sampling\ rate]$ multiplied by $[frequency\ resolution]$. The choice of frequency resolution is a trade-off between noisy and biased estimation. Further discussion on the statistics of estimation is available in Section 9.2 of Bendat and Piersol[7].

Measurements are discrete. Measurements made at a finite rate with finite separation intervals form a finite set that cannot contain all the information of a continuous signal. Diffraction phenomena arise from interactions of the separation interval of the samples and frequency components of the signal. These *aliasing phenomena imply that a single set of samples could come from many different signals of different frequency*. The different frequencies of so-called 'aliased signals' produce the same result after sampling even though they are different before sampling. *One frequency becomes the alias of another after they are sampled.* The different *sampled signals* are aliases of each other although the original *Un*sampled signals are independent, and not aliases of each other at all. The aliases do not resemble each other, even qualitatively, and so studying the aliases can be highly misleading if their existence is not understood, or their importance is discounted.

Because measurements are of finite duration, the sampling process is truncated in time. Truncation produces discontinuities in functions *and their time derivatives* that create errors beyond the simple limitations in resolution produced by the finite time of measurement (the Gibbs phenomenon rediscovered by and named after the 19th century Connecticut (USA) physical chemist J.W. Gibbs). These wide-ranging errors arise because the Gibbs phenomena produces *nonuniform convergence* in the various integrals that estimate frequency domain functions. Windowing methods *are required by the mathematics of truncation* (and nonuniform convergence) to avoid serious artifacts throughout the frequency domain. An optimal solution to these issues is known[62] but nonoptimal tradeoffs between overshoot and accuracy are often adopted when they are well-suited to particular applications.

**Driven Damped Spring**. To demonstrate estimation of the coherence and frequency response function, we consider a system in which we can determine these functions analytically: an object with mass $m$ (right hand side) attached to a piston (left hand side) by a damped spring (Fig. 2). The object is subjected to a driving force that prevents it from coming to rest. In this case, the frequency response function is well-known in the physics and engineering literature and is reproduced here to introduce notation and for completeness.

+–-

Figure 2

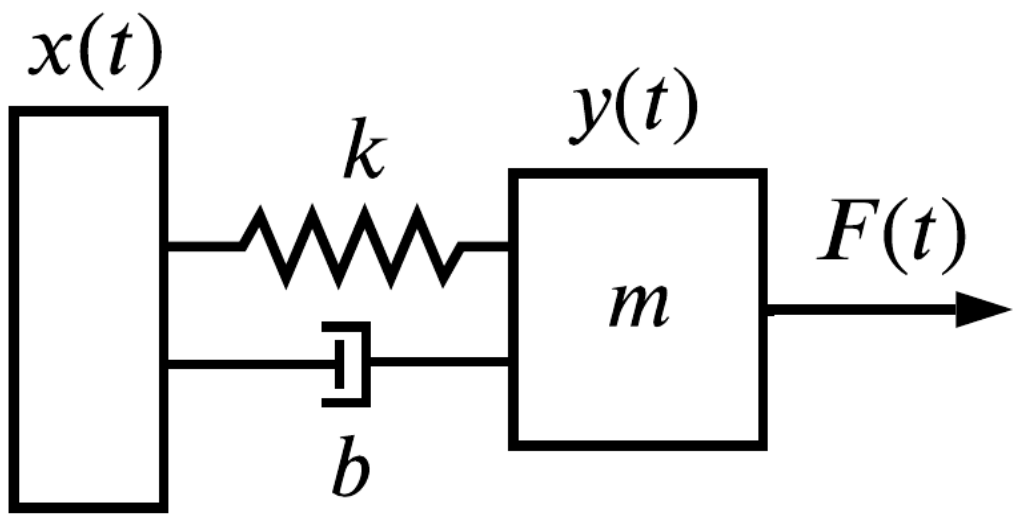

*Figure 2. Schematic of a damped spring* with spring constant $k$ in N m$^{-1}$, mass $m$ in kg, and damping $b$ in N S m$^{-1}$, driven by the force $F(t)$ (N).

Let us focus on the situation where the force $F(t)$ is a result of driving the position of the piston. The position of the piston is the input $x(t)$, applying a force $F(t) = kx(t)$ onto the mass. The Fourier transform of the input is $X(f)$. The output is the location of the mass, $y(t)$ or $Y(f)$. We also define the displacement $z = y - x$. The driven damped oscillator is described in many sources including Ch. 5.5 of Taylor[63], which we depend on for definitions, analysis, and discussion, except that Taylor uses a different coordinate system and does not specify the piston as the source of the driving force. Taylor is particularly helpful in explaining the frequency domain in physical language with minimal mathematics and no use of Laplace transforms. The classical Laplace transform approach is found throughout the engineering literature, for example, in the digital context in [61].

The equation of motion for a system with mass $m$ and damping $b(dy/dt)$ and damping constant $b$ driven by forcing function $F(t)$ is derived from Newton's law to be,

$$m\frac{d^2y}{dt^2} + b\frac{dy}{dt} + ky = F(t). \tag{11}$$

It is customary to use reduced variables, one set of which are,

$$\omega_0 = \sqrt{\frac{k}{m}} \tag{12}$$

$$\beta = \frac{1}{2}\frac{b}{m} \tag{13}$$

$$f(t) = \frac{F(t)}{m}. \tag{14}$$

The standard form is then,

$$\frac{d^2y}{dt^2} + 2\beta\frac{dy}{dt} + \omega_0^2 y = f(t). \tag{15}$$

This system is customarily analyzed with a time-dependent sinusoidal force with angular frequency $\omega$ that drives the mass described by either $f(t)$ or $F(t)$,

$$f(t) = f_0 \cos(\omega t), \tag{16}$$

or $F(t) = mf_0 \cos(\omega t)$. Note that $f_0$ is not a function of time.

The steady-state response (after transients damp into nothing) is

$$y(t) = f_0 \cos(\omega t - \delta), \tag{17}$$

with amplitude $A(\omega)$,

$$A = \frac{f_0}{\sqrt{(\omega_0^2 - \omega^2)^2 + 4\beta^2\omega^2}}, \tag{18}$$

and phase angle $\delta(\omega)$,

$$\delta = \arctan\frac{2\beta\omega}{\omega_0^2 - \omega^2}. \tag{19}$$

Taylor[63] (Section 5.8) has a useful discussion of the time domain response of the system and its connection with the frequency functions.

## Methods

**MD simulation.** We used code developed in the Minh group (https://github.com/swillow/pdb2amber) to prepare a model of the system. A crystallographic structure of crambin (Protein Data Bank 1CRN) was protonated using PROPKA[64] at pH 7. The protonated protein was inserted into a cubic box of water with 0.1 M NaCl. The AMBER ff14SB force field[65] was used for the protein, OPC3 parameters[66] for water, and Joung and Cheatam TIP4P/EW parameters for ions [67]. The length of each side of the water box was 75 Å, much larger than crambin. Preliminary calculations with a smaller box of 45 Å showed that crossing the boundary affects the coherence. To avoid the ambiguity of unwrapping, we only report results from the system in the larger box in which *the crambin molecule never crossed the periodic boundary*.

MD simulation was performed with OpenMM version 7.4.2[68]. First, the system was minimized. Isothermal MD was performed using the Langevin integrator at temperature $T = 300\ K$ with a time step of 2 femtoseconds (fs) for 100 ns. (Shorter simulations performed with the deterministic Verlet integrator gave indistinguishable results). Samples were stored every 0.01 nanoseconds (ns) (equivalent to a rate of 100 GHz).

**Estimation:** One technical issue with linear systems analysis of molecular configurations is that atoms are specified in three dimensions but the methods are designed for one dimension. In our analyses of MD simulations, we circumvented this issue by reducing the dimensionality of the signals. For the input and output signals, we used the mean-normalized displacement of the atom or the mean position of the atom group,

$$d(t) = \sqrt{x^2(t) + y^2(t) + z^2(t)} - \langle \sqrt{x^2(t) + y^2(t) + z^2(t)} \rangle \qquad (20)$$

where $x(t)$, $y(t)$, and $z(t)$ are the coordinates of the atom (group) and the brackets denote an average over the time series.

For signals from each atom pair, we used the classical Welch periodogram method to estimate the energy at each frequency[8]. The **pwelch** method described on pg. 6-21 of the MATLAB Signal Processing Toolkit User's Manual[1] was used to calculate $\hat{G}_{xx}(f)$ and $\hat{G}_{yy}(f)$. For most trajectories analyzed, 32 disjoint Hann(ing) windows were used. The number of DFT points was the highest power of two less than the ratio of the total number of samples and number of disjoint windows. In pseudocode, this is,

$$\text{Number of DFT points} = 2^{\text{FLOOR}\left(\log_2 \frac{\text{number of time steps}}{32}\right)}.$$

The percent overlap between windows was always kept to 50%. The bias-variance tradeoffs from percentage overlap and number of disjoint windows has been discussed in detail elsewhere[7, 8]. Similar parameter estimation in the **cpsd**[1] (pg. 6-28 of the User's Manual[1]) gives us $\hat{G}_{xy}(f)$. Frequency response and coherence functions were calculated from power spectra using Eqs. (7) and (8).

# Results

**The frequency response and coherence function can be accurately estimated for the driven damped spring.** We performed simulations of the driven damped spring and used the time series of inputs and outputs to estimate the frequency response and coherence function (Fig. 3). The estimated coherence is indistinguishable from one. Thus, coherence analysis of the input and output time series correctly shows that the system is linear. Moreover, for the two sets of parameters that were tested, the estimated frequency response function is indistinguishable from the analytical expression, Eq. (18). These results demonstrate that the methods we have used can accurately estimate the coherence and frequency response function without significant systematic error.

Figure 3

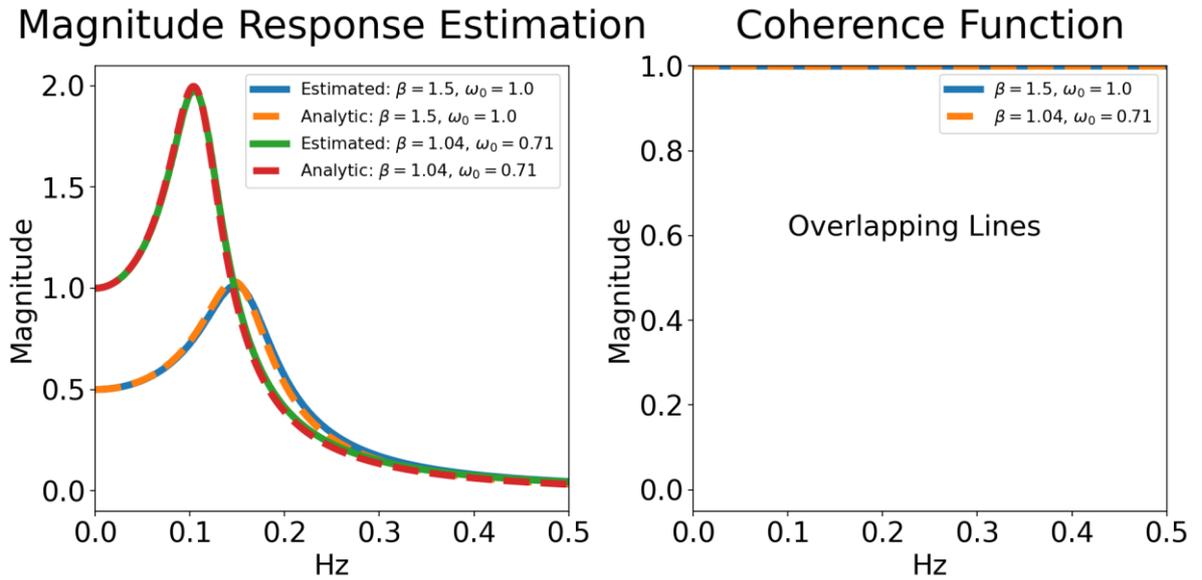

*Figure 3. Magnitude of the frequency response (left) and coherence function (right) for the driven damped harmonic oscillator.* The analytical amplitude $A(\omega)$ is from Eq. (18). Simulations were performed using the 'lsim' command[2] in MATLAB. 65536 samples were collected at a rate of 2 Hz. Power spectra are shown in Fig. 4 and computed as described in the corresponding caption. The estimated magnitude of the frequency response function $\hat{A}(f) = |\hat{H}(f)|$ and coherence function was based on the ratio of power spectra $\hat{G}_{xx}(f)$ and $\hat{G}_{yy}(f)$ and cross power $\hat{G}_{xy}(f)$ of the frequency function, Eq. (7).

In contrast to the smooth estimate of the frequency response and coherence function, the power spectra appear to be remarkably noisy (Fig. 4). The apparent 'noisiness' of signals derived from stochastic time series (like $x(t)$ and $y(t)$) can be misleading. It is a sensitive function of the number of points that happen to be shown in the figure. The number of points that happen to be used in Fig. 4 is 256. Explicit formulas for variance, bias, and confidence limits are given by Bendat and Piersol[7] and also Table S-1.

Figure 4

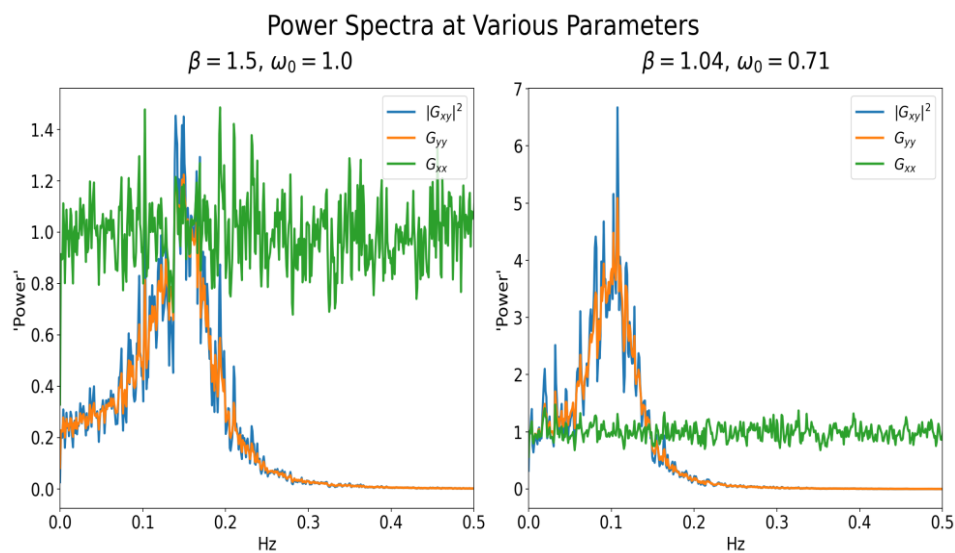

*Figure 4. **Power spectra*** used for estimating the frequency and coherence functions in Fig. 3. Power spectra were calculated using Welch's method implemented in the MATLAB Signal Processing Toolbox[1] as 'pwelch', with a Hanning window size of 16 points and 50% overlap between windows. Cross power was computed using 'cpsd' with the same parameters.

**H-bonds in the α-helices of Crambin are linear systems.** After performing a linear systems analysis of a damped spring, we moved on to analyzing molecular dynamics simulations of crambin. We first considered atoms interacting via H-bonds in the α helices (Fig. 5).

For all the H-bonds in α-helices of crambin, the coherence for frequencies below 10 GHz (about 0.1 ns) is close to one (Fig. 6). For the main α helix, the mean coherence for all frequencies between 0.391 GHz and 5.08 GHz (corresponding to reciprocal times of 2.6 ns and 0.2 ns, respectively), is 0.94 or above for all but one pair (Table 1); the mean coherence between LEU18N and VAL15O is slightly lower, 0.919. Comparable results are seen for the small α helix (Table 1). Thus, nearly all the low-frequency motion of the oxygen on one end of the H-bond is determined by the motion of the nitrogen on the other end.

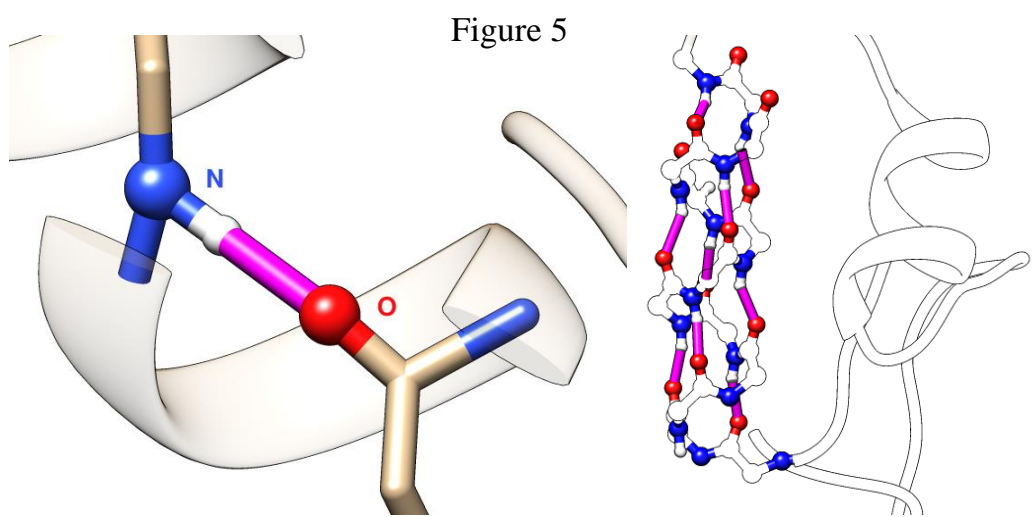

Figure 5

*Figure 5. **H-bonds in the main α-helix of crambin.*** A high-resolution view of a single H-bond (left) and medium-resolution view of all the H-bonds in the main α-helix (right). The right panel also shows the secondary structure of the smaller α-helix.

Figure 6

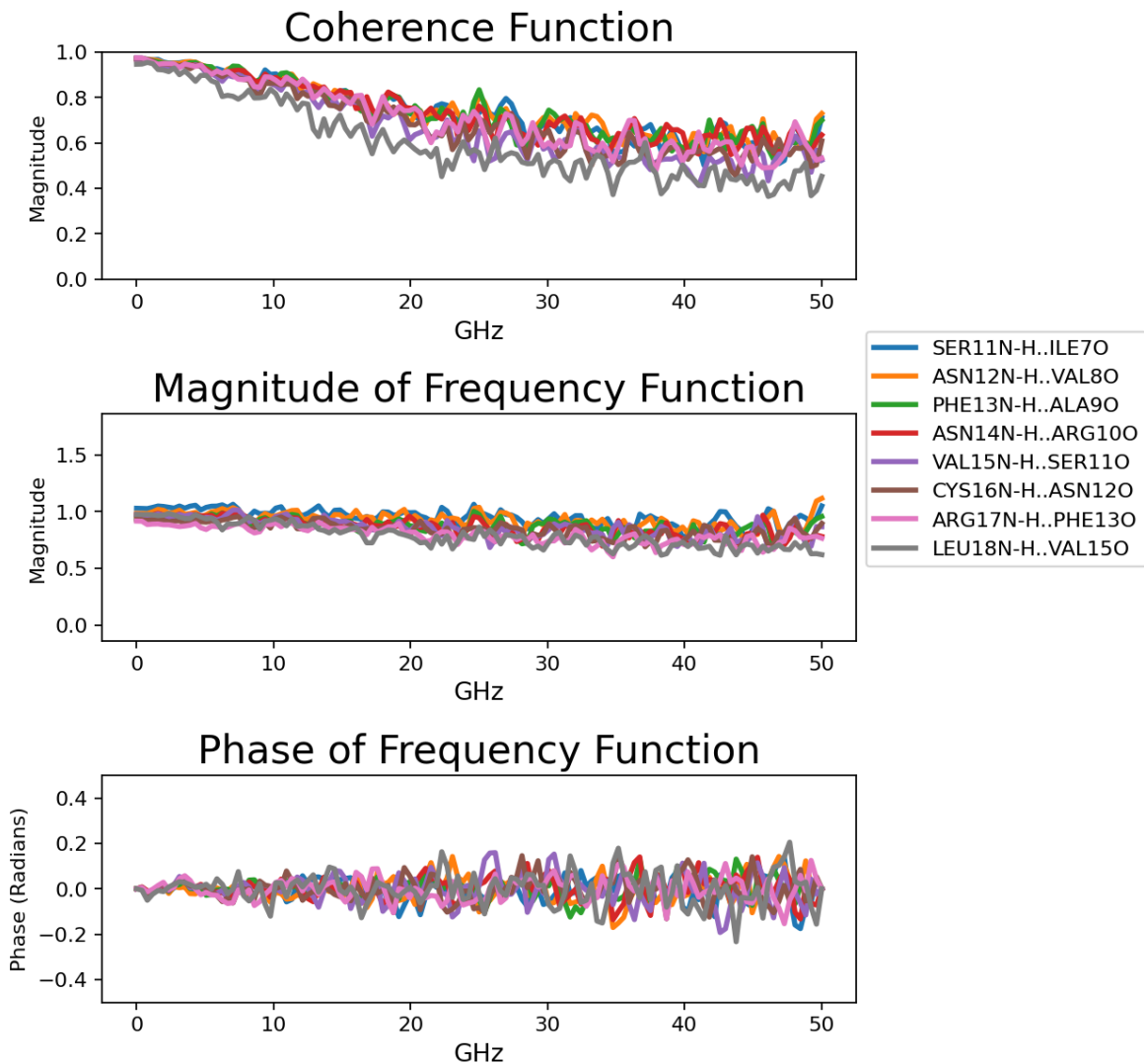

*Figure 6. Estimated coherence and frequency response functions* for H-bonds in the main α-helix of crambin.

In several cases, we have also verified through swapping input and output functions what is known through mathematical analysis: the coherence $C_{xy}(f)$ between $x$ and $y$ is equal to the coherence $C_{yx}(f)$ between $y$ and $x$. The frequency functions $H_{xy}(f)$ and $H_{yx}(f)$ are also equivalent. Therefore, the statement about motion of the atom pair may be reversed: nearly all the low-frequency motion of nitrogen in the H-bond is determined by the motion of oxygen on the other end of the H-bond and *vice versa*. Thus, **when averaged over at least 0.1 ns, the H-bonds move like springs.**

On the other hand, for higher frequencies (shorter times) the coherence function is significantly smaller than one. For these frequencies, the H-bonds do not appear to be a linear system; motion of the atoms over a time scale less than 0.1 ns cannot be accurately described with a harmonic oscillator model. It is possible that nonlinearity reduces the value of coherence. Another possibility is that the atoms have significantly different inputs of power from other atoms and structures. The coherence function shows the length of time that needs to be averaged over for a linear system to accurately model the MD.

| Table 1 Coherence and frequency response function for the α-helices, averaged over frequencies between 0.391 GHz and 5.08 GHz | | | | |
|---|---|---|---|---|
| **Donor (Input)** | **Acceptor (Output)** | **Mean Coherence** | **Mean Magnitude** | **Mean Phase** |
| Main α-helix | | | | |
| SER11N | ILE7O | 0.959 | 1.038 | -0.001 |
| ASN12N | VAL8O | 0.957 | 0.992 | -0.004 |
| PHE13N | ALA9O | 0.953 | 0.955 | 0.006 |
| ASN14N | ARG10O | 0.947 | 0.966 | 0.010 |
| VAL15N | SER11O | 0.944 | 0.964 | 0.010 |
| CYS16N | ASN12O | 0.948 | 0.920 | 0.008 |
| ARG17N | PHE13O | 0.950 | 0.884 | 0.012 |
| LEU18N | VAL15O | 0.919 | 0.964 | 0.001 |
| Small α-helix | | | | |
| ALA27N | GLU23O | 0.938 | 1.000 | -0.006 |
| THR28N | ALA24O | 0.915 | 1.012 | -0.016 |
| TYR29N | ILE25O | 0.937 | 0.970 | -0.007 |

The frequency response function is consistent with this H-bond behaving approximately as a spring with no noticeable damping. At low frequencies, the magnitude is flat and near one. For a spring, a flat magnitude near one is observed at frequencies significantly smaller than the natural frequency. Moreover, the phase is near zero at low frequencies. A phase of zero is expected in a spring without damping. The behavior at higher frequencies is less clear; both the magnitude and phase of the frequency function appear to be noisier. Because the coherence is lower and system is not linear at these frequencies, the estimated frequency response function is no longer meaningful and not interpretable in terms of an analytical model like the spring.

**Non-interacting atoms have negligible coherence.** As a baseline for comparison, we also consider the coherence of atom pairs that are not expected to interact. Fig. 7 shows the coherence for atoms that are physically distant and separated by the length of many covalent bonds. In contrast to the H-bonds, the coherence is near zero. We have checked many such pairs and they all give similar results. Unrelated atoms do not share power. Their motions are uncorrelated. A frequency function between them is meaningless and so it is not shown. The low coherence observed in the noninteracting pair suggests that the high coherence observed in the α helices is not simply an artifact of the estimation procedure.

Figure 7

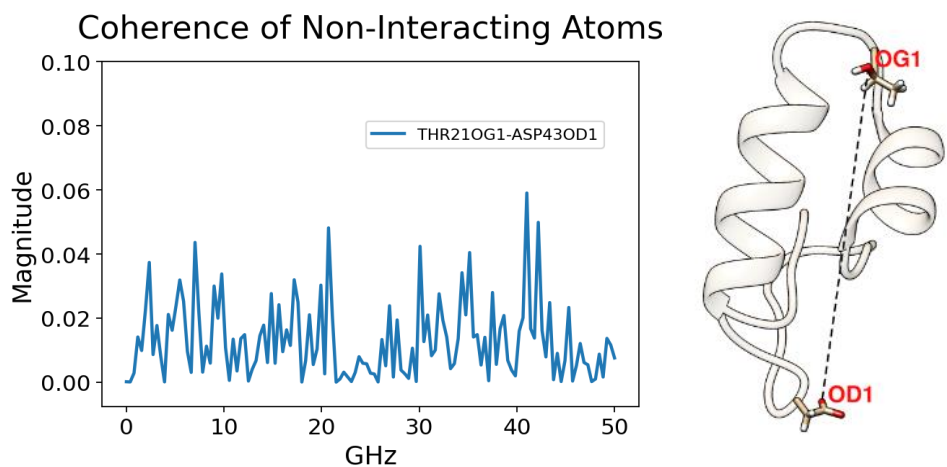

*Figure 7: Coherence function between two unrelated atoms,* OG1 on THR21 and OD1 on ASP43 (left). These atoms were not expected to have any interaction because their separation is about

**Coherence is insensitive to translation and rotation**. As additional controls, we evaluated the effect of translating and rotating the system. Although the following results were obtained for PHE13N and ALA9O, we observed comparable results for all tested atom pairs.

First, we tested translation. We performed 100 calculations in which we selected a random displacement between -200 and 200 Å and applied it to every atom in every frame of the simulation. Estimated coherence functions for every translation are shown in Fig. S-1. We observe that at low frequency, translation leads to very small differences between the estimated coherence values (Fig. S-1). Translations do not affect the result that the coherence at these frequencies is greater than 0.9. Over the 100 translations, the mean coherence between 0.39 and 5.09 GHz has an average value near 0.95 and a standard deviation of only $4.69 \times 10^{-3}$ (Fig. S-2)! At higher frequencies, in contrast, the mean coherence itself is lower and there are larger variations in the estimated coherence function across the different translations.

We also tested rotation. The coherence before and after applying a random rotation is indistinguishable (Fig. S-3).

**Coherence is sensitive to wrapping**. Historically, MD became feasible in atomic detail[69] when it adopted periodic boundary conditions that dramatically decreased the amount of computation required. However, as discussed by many authors including p. [31]35-38 of [31], periodic systems are not the same as non-periodic systems; a price was paid for feasibility. One cost is that when atoms leave one side of the box, they enter on the other, producing an artificial discontinuity. The actual continuous paths of atoms are artifactually wrapped by the periodic boundary conditions of MD into discontinuous paths. The discontinuity in paths produces artificial correlations and may introduce significant errors[7, 8, 32, 33, 34, 70, 71].

We observed that the coherence function for atoms that have a discontinuous path due to periodic boundary conditions is greater than 0.99. In other words, wrapped trajectories had artificially large coherence. We did not use these calculations to produce the figures and tables in this work.

We thought of and tried two ways to address the issue of discontinuous paths:

(1) Unwrapping trajectories. We used PBCTools in the VMD[72] software package to partially remove the artifacts of wrapping[70, 71]. Any trajectory that became discontinuous due to some part of the protein crossing the boundary was made continuous.

(2) Performing simulations in a box and under conditions so that crambin does not reach an edge of the box and so has continuous trajectories.

We first used unwrapping. Out of an abundance of caution, we also repeated the calculations with a larger box and confirmed that crambin does not cross a box edge. Fortunately, although theory suggests that the unwrapping procedure is inexact[70], in the cases we tried the standard unwrapping procedure gave similar results as the large box. This suggests that is it reasonable to estimate coherence for trajectories that have been made continuous by the standard unwrapping procedure.

**Other atom pairs have high coherence.** In addition to H-bonds in the α-helices, we also calculated the coherence of other atom pairs that we thought could have high coherence: covalent bonds, H-bonds in the β-sheet, and salt bridges.

For covalent bonds, coherence is expected to be high one because the molecular mechanics force field specifically includes a harmonic potential between the atoms. The coherence may be reduced by the contribution of energy terms and other atoms. Indeed, we find high coherence between atoms connected by covalent bonds. For majority of pairs involving the alpha and its neighboring carboxyl carbon, the coherence is near one at low frequencies and 0.9 or above for even the highest frequencies (Fig. 8). For peptide bonds, the coherence is higher than for H-bonds, but not quite as high as between alpha and carboxyl carbons.

Figure 8

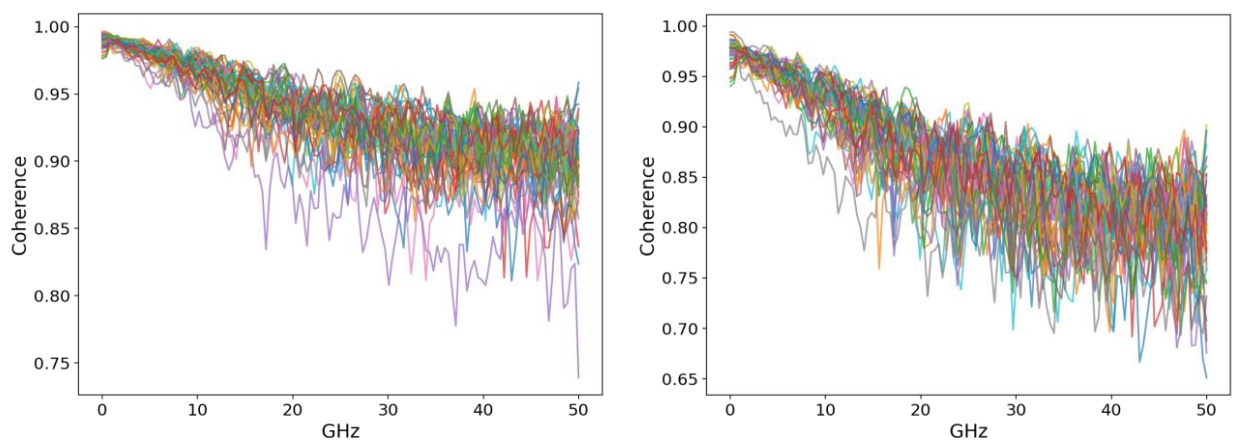

*Figure 8: Coherence function for atoms across covalent bonds in crambin* for alpha and carboxyl carbons (left) and carbon and nitrogen atoms across peptide bonds (right). Note the range on the y axis.

It seemed worthwhile to investigate the coherence of the other H-bonds in crambin, particularly those in the β-sheet portion of the protein. Table 2 identifies the residues analyzed (residues 1 through 4 and 32 through 35) and shows that the H-bonds of β-sheets and α-helices are much the same; we obtain similar results to that of the α-helices documented in Table 1.

**Table 2**
**Beta Strand**

| Donor (Input) | Acceptor (Output) | Mean Coherence | Mean Magnitude | Mean Phase |
|---|---|---|---|---|
| ILE35N | THR1O | 0.922 | 0.922 | -0.003 |
| ILE33N | CYS3O | 0.933 | 0.947 | -0.004 |
| CYS3N | ILE33O | 0.930 | 0.986 | 0.004 |

We also investigated the coherence in a salt bridge[24, 26], imagining that the estimated coherence might be useful in evaluating putative salt bridges in less well-defined systems. The salt bridge studied was between ASN46 OXT…ARG10 NE and ASN46 O…ARG10 NH2 as defined in [26]. The mean coherences between the two were 0.906 and 0.931 respectively with frequency function estimates being like previously shown estimates.

**High coherence is observed between atom groups**. We thought it would be interesting to explore whether there is coherence between groups of atoms. We performed a preliminary analysis based on leaves of the main α-helix, which consists of residues 7 through 19 (Fig. 9). We defined a leaf as the first alpha carbon to the third alpha carbon (exclusive) with all backbone atoms in between. Other definitions of leaves are possible and may be superior. We calculated the coherence between adjacent leaves. The trajectories of the eight atoms in each leaf were averaged to define the position of each leaf. The average displacement of the leaves was used as input and output.

Interactions between leaves have similar behavior to the interactions between atoms in the H-bond. Each leaf has high coherence and appears to interact as a linear system with the next leaf (Table 3). The frequency response function has a magnitude near one and phase near zero. These preliminary results suggest that coherence can help identify groups of atoms that move together as a protein changes conformation and interact as linear systems.

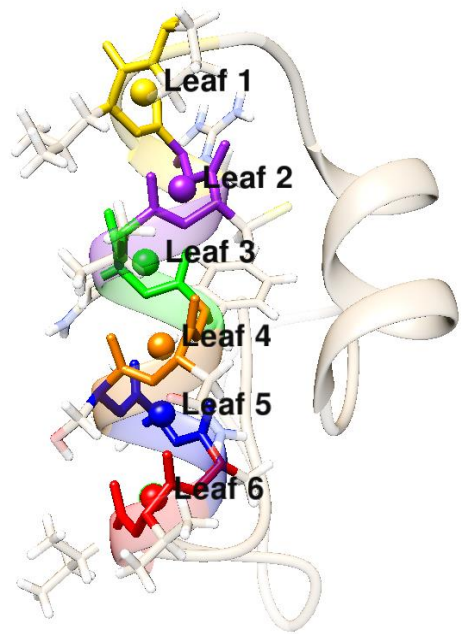

*Figure 9:* Leaves in the main α helix of crambin.

| Table 3. Coherence of leaves of the main α-helix of crambin | | | |
|---|---|---|---|
| Input and Output | Mean Coherence | Mean Magnitude | Mean Phase |
| Leaf 1 and 2 | 0.948 | 0.865 | 0.002 |
| Leaf 2 and 3 | 0.938 | 1.044 | 0.003 |
| Leaf 3 and 4 | 0.929 | 0.938 | 0.006 |
| Leaf 4 and 5 | 0.936 | 1.070 | 0.010 |
| Leaf 5 and 6 | 0.918 | 1.058 | 0.001 |

# Discussion

**Linear systems analysis has a broad scope.** We have used classical methods developed to analyze stochastic time series[4] in engineering to analyze MD trajectories. The methods do not depend on the properties of, or the existence of a statistical probability distribution, e.g. the Boltzmann distribution. In this way, they are even more robust than nonparametric methods[30]. They do not require models of the interactions other than the force field inherent in MD itself. The analysis only depends on the locations of atoms computed in MD simulations. It only requires the existence of the Fourier transforms of the atomic motion and has no other assumptions.

One potential implication of the broad scope of linear systems analysis is the application to MD of systems out of equilibrium. While MD of biomolecules is usually performed under assumptions of equilibrium, many biomolecules – including ion channels, transporters, and many enzymes – can be thought of as devices that primarily operate out of equilibrium. In the same way that electromechanical devices use a power supply to transform an input signal or multiple input signals into an output signal, these biomolecular systems use concentration gradients, voltages, and high-energy molecules to drive the flows of ions and the production of chemical species. The distribution of configurations and kinetics of transitions between them are likely to be altered by changes to input signals and flows of ions, substrates, and products. Hence, the most functionally relevant MD simulations may not be of the biomolecules in equilibrium, but of relaxation processes and nonequilibrium steady states. In the same way that engineers have turned to linear systems analysis for the investigation of electromechanical devices, the biomolecular MD community can employ the approach to evaluate the most functionally relevant MD simulations of biological devices.

**There is precedent for using linear systems analysis in MD.** We are not the first to apply this class of engineering methods to MD. The power spectra of individual trajectories in MD give important information as so well stated in the key papers that serve as a portal to this literature[5, 21], e.g., on p. 2019 of [21] explains that "Power spectra contain all peaks that can be found in the IR spectrum, and even more – they can be imagined as the sum of the IR spectrum, the Raman spectrum and all motions that are neither IR nor Raman active." We have extended these powerful results motivated by an analogy between the motions of two atoms in a 'chemical' bond and the motions of charges at the input and output of electronic devices, typically a linear system (in the engineering sense defined previously in this paper) like a resistor, capacitor, or amplifier.

**Many atom pairs are linear systems.** Specifically, we have estimated the coherence between the displacement of atom pairs in plant protein crambin with different types of interactions —- H-bonds, covalent bonds, salt bridges, and no apparent interactions — to see if they constitute an engineer's linear system. If an analysis shows that the motions of two objects share the same power (i.e., are so correlated that their coherence is one), the objects can be described by a linear system (in the engineering sense defined previously). If they are not so correlated, no conclusions can be drawn. Indeed, it is important not to analyze the motions in that case, without changing the underlying model. To our surprise and delight, in many cases, we found that many atom pairs are linear systems! In these cases, most of the (power of the) motions of one atom are linearly related to the (power of the) motions of the other atom, and *vice versa*.

We did not find all pairwise atomic interactions to be linear. For interactions in which the coherence is not near one, an extension of this work may involve exploring multiple-input-multiple/single-output (MIMO/MISO) systems. This approach is needed if the power on the two atoms (of the coherence function) is not shared, because components of power are added to one atom and not the other.

**<u>Coherence may help simplify interatomic potentials.</u>** Identification of atomic interactions as linear systems is a delight because it describes all the mechanical motions of the two atoms that are analyzed. If the motion of one atom is known, the motion of the other atom is predicted by a <u>simple spring model</u> with a handful of parameters. These parameters of the linear system model are robust. They do not change with conditions. If the system is linear, all its motions (in the time and frequency domain) are predicted by the model. The motion of one determines the motion of the other whether the motion is described in the time or frequency domain. Moreover, since we found that there is no phase delay, ***the interatomic potential for these interactions is accurately modeled as a simple harmonic oscillator***.

One reason that the finding that H-bonds and salt bridges can be accurately modeled as a simple harmonic oscillator is surprising because the molecular mechanics force field used to produce the trajectory represent their interactions with more complex functions. The AMBER force field uses the Lennard-Jones potential for steric and Coulomb potential for electrostatic interactions. However, the structure of the protein forces these atom pairs to act as if they were connected by a simple harmonic oscillator.

While it is a delight, the finding of a linear system is also a challenge because it does not establish how to estimate its parameters. While there appears to be no damping, the spring constant cannot be estimated without further assumptions and analysis. A frequency at which high coherence is observed is not the natural (characteristic) frequency of the harmonic oscillator. If a resonance peak were clearly visible in the frequency response functions, then the spring constant could potentially be inferred by fitting the analytical model to the observed function. Unfortunately, the peak is not observed. There are several other possible routes to spring constants, of which we will mention two.

One possible route to an approximate spring constant is to assume that populations are given by the Boltzmann distribution and that the spring is the only force (i.e., that it is the only term in the potential force fields of MD) experienced by the two atoms. In this case, the energy as a function of the distance $z$ is given by, $U(z) = \frac{1}{2}k(z-\bar{z})^2$, where $\bar{z}$ is the equilibrium distance. Assuming the Boltzmann distribution gives the probability density proportional to $\exp\left[-\frac{k}{2k_BT}(z-\bar{z})^2\right]$, where $k_B$ is Boltzmann's constant and $T$ is the temperature. Thus, if $\hat{\sigma}$ is the estimated standard deviation of the displacement, $\hat{k} = k_BT/\hat{\sigma}^2$ is an estimate of the spring constant.

Another way to define spring constants is to minimize the energy of the system and then compute the second derivative of the potential energy with respect to the interatomic distance. Use of the second derivative in this way is analogous to normal modes analysis[73], where eigenvectors and eigenvalues of the Hessian matrix with respect to atomic positions are used to define normal modes and frequencies. However, this approach has a drawback if there is roughness in the energy landscape: the second derivative may describe local opposed to global curvature.

**Linear systems analysis has connections with but has important distinctions from normal modes analysis**[73, 74]. A few more words on normal modes analysis are in order. Normal modes analysis is based on the second derivative of the potential energy with respect to the coordinates of every atom. It assumes that the potential energy surface of the molecule is described by a multidimensional harmonic oscillator. On the other hand, the coherence analysis in this paper assesses whether the interatomic potential between a pair of atoms is accurately described by a spring. Our conjecture is that high coherence is a sufficient but not necessary condition for structural fluctuations to be described by normal modes analysis. That is, if there is high coherence between every pair of atoms in a group, then the interatomic potential energy functions comprise coupled harmonic oscillators, as in an elastic network model[75]. The dynamics of such a system can be accurately described by normal modes. On the other hand, it is possible that in some systems normal modes may still be a good approximation even if not all atom pairs in a group are coherent. The connection between these two techniques should be explored more thoroughly in future work.

**Coherence may help define noncovalent interactions and secondary structure**. The finding that coherence is high for many atom pairs with known interactions but negligible for others with no apparent interactions suggests that coherence may be used to define noncovalent interactions and secondary structure. Purely geometric considerations for noncovalent interactions involve subjective thresholds, e.g. the lengths and angles of H-bonds. For this reason, the IUPAC definition of the H-bond[76] notes the existence of "borderline cases for which the interpretation of the evidence might be subjective" and that "new criteria for hydrogen bonding could evolve" with further progress in experimental and theoretical methods. High coherence may be one of these new criteria. High coherence may be appropriate criterion not only for defining H-bonds, but also salt bridges, hydrophobic attractions, and van der Waals interactions. Beyond these known interactions, coherence analysis may help identify strong interactions in proteins that have not previously been catalogued. Finally, they could also be helpful in defining secondary structure.

To be specific, let us consider an example from crambin. While some visualization software that we tried show LEU18 and PRO19, located at the end of the main α-helix, as part of the main α-helix, others do not; geometric considerations used by various software may be ambiguous. However, the fact that coherence of a H-bond involving LEU18 is quantitatively lower, at 0.919, suggests that it has less α-helical character than other parts of the helix.

**Coherence may help develop reduced models, helping define rigid bodies and the interactions between them**. MD simulations and cryo-electron microscopy (cryo-EM) show that protein motion can often be described as the relative translation and rotation of rigid bodies connected by flexible linkers. This has inspired one of us (Minh) to use rigid body simulation methods from robotics to study biomolecular motion.[77] As every pair of atoms within a rigid body has high coherence, coherence may be applied to an atomistic MD simulation to identify atoms that are suitable to group into a rigid body. Specific ways to perform this grouping will be the subject of future research. Moreover, as hinted in our analysis of leaves in the main α-helix, the coherence function may also be useful in investigating interactions between atom groups including rigid bodies. If pairs of atom groups interact linearly, then their motion may be described by constant coefficient linear differential equations. Otherwise, reduced models[78] may require more complex potential energy functions. For example, it has been suggested that united residue models require multi-torsion terms, e.g. based on self-consistent cumulant expansions[79], to account for correlation between virtual-bond dihedral angles along straight chain segments.

In either case, reduced models can be tested by experiments that vary the composition of atom groups or modify their motion by known interventions such as ligand binding or altering the composition and concentrations of ions. If the reduced models help understand such a range of protein behavior, they will be a remarkable help in linking protein structure, dynamics, and function.

**Coherence analysis should be applied to more systems**. Given our intriguing results for crambin, it is worth applying coherence analysis to MD simulations of other proteins and other systems. As noted, crambin is particularly rigid. However, coherence may be especially helpful in the context of more flexible proteins with larger conformational variability, where it may be easy to overlook the importance of correlations imposed by macromolecular structure. Beyond proteins, the coherence of the Oxygen-Oxygen H-bond of liquid water and ice would be an interesting comparison. In the same way that protein structure stabilizes the H-bonds in α-helices, we expect that ice provides a macroscopic structure – scaffold if you wish – that orients the orbitals of oxygen, hydrogen, and oxygen so that the 'internal' chemical energy of interaction can better compete with entropic randomness, stabilizing the H-bonds. This structural effect may be evident from a comparison of the coherence function.

# Conclusions

Coherence was used to analyze the motion of pairs of atoms in molecular dynamics simulations of crambin. High coherence was observed for low-frequency motion (below 10 GHz) between many pairs of atoms that we anticipated to have strong interactions: covalent bonds between backbone atoms, H-bonds within alpha helices and beta sheets, and salt bridges. Coherence was lower for lower-frequency motion. It was negligible for atom pairs that are physically distant and separated by many covalent bonds. For high coherence atom pairs and frequencies, the frequency response function was indicative of an undamped spring. These results suggest that coherence could become a useful tool for the analysis of molecular dynamics simulations, with potential applications in identifying interactions, developing reduced models, and investigating allostery.

# Acknowledgement

This research was catalyzed by funding from the Illinois Institute of Technology. Stanley Nicholson is a joint (coterminal B.S./M.S.) student with tuition funded in full by a Camras Scholarship. Costs were generously supported by the Robert E. Frey, Jr. Endowed Chair in Chemistry held and administered by David Minh.

# Supporting Information

Provided in a supplementary document:

1) Estimates of Random Error are quoted from the literature
2) Systematic Errors of Translation and Rotation are evaluated/

# Authors' Information

[*]Robert Eisenberg
bob.eisenberg@gmail.com
Department of Applied Mathematics
Illinois Institute of Technology
Chicago IL

Stanley Nicholson

*Email address:* snicholson1@hawk.iit.edu

David Minh
*Email address:* dminh@iit.edu

---

[*]*Corresponding Author*

# References

(1) MathWorks. *Signal Processing Toolbox, Version 7.5, Release 2017b*; 2017.

(2) MathWorks. *Control System Toolbox Reference, R2022a, Version 10.11.1*; 2017.

(3) Fiorin, G.; Klein, M. L.; Hénin, J. Using collective variables to drive molecular dynamics simulations. *Molecular Physics* **2013**, *111* (22-23), 3345-3362. Tubiana, T.; Carvaillo, J.-C.; Boulard, Y.; Bressanelli, S. TTClust: a versatile molecular simulation trajectory clustering program with graphical summaries. *Journal of chemical information and modeling* **2018**, *58* (11), 2178-2182.

(4) Avval, T. G.; Moeini, B.; Carver, V.; Fairley, N.; Smith, E. F.; Baltrusaitis, J.; Fernandez, V.; Tyler, B. J.; Gallagher, N.; Linford, M. R. The Often-Overlooked Power of Summary Statistics in Exploratory Data Analysis: Comparison of Pattern Recognition Entropy (PRE) to Other Summary Statistics and Introduction of Divided Spectrum-PRE (DS-PRE). *Journal of chemical information and modeling* **2021**, *61* (9), 4173-4189. Yasuda, T.; Shigeta, Y.; Harada, R. Efficient Conformational Sampling of Collective Motions of Proteins with Principal Component Analysis-Based Parallel Cascade Selection Molecular Dynamics. *Journal of chemical information and modeling* **2020**, *60* (8), 4021-4029. Ramanathan, A.; Agarwal, P. K.; Kurnikova, M.; Langmead, C. J. An online approach for mining collective behaviors from molecular dynamics simulations. *J Comput Biol* **2010**, *17* (3), 309-324. DOI: 10.1089/cmb.2009.0167. Stoltz, G. Error estimates and variance reduction for nonequilibrium stochastic dynamics. **2022**. DOI: 10.48550/arxiv.2211.10717.

(5) Thomas, M.; Brehm, M.; Fligg, R.; Vöhringer, P.; Kirchner, B. Computing vibrational spectra from ab initio molecular dynamics. *Physical Chemistry Chemical Physics* **2013**, *15* (18), 6608-6622.

(6) Frömbgen, T.; Blasius, J.; Alizadeh, V.; Chaumont, A.; Brehm, M.; Kirchner, B. Cluster Analysis in Liquids: A Novel Tool in TRAVIS. *Journal of chemical information and modeling* **2022**.

(7) Bendat, J. S.; Piersol, A. G. *Random data: analysis and measurement procedures Fourth Edition*; John Wiley & Sons, 2010.

(8) Carter, G.; Knapp, C.; Nuttall, A. Estimation of the magnitude-squared coherence function via overlapped fast Fourier transform processing. *IEEE transactions on audio and electroacoustics* **1973**, *21* (4), 337-344.

(9) Dodds, C. J.; Robson, J. D. Partial coherence in multivariate random processes. *Journal of Sound and Vibration* **1975**, *42* (2), 243-249. DOI: 10.1016/0022-460x(75)90219-9.

(10) Cadzow, J.; Solomon, O. Linear modeling and the coherence function. *IEEE transactions on acoustics, speech, and signal processing* **1987**, *35* (1), 19-28.

(11) Thomas, C. W. Coherence function in noisy linear system. *International Journal of Biomedical Science and Engineering* **2015**, *3* (2), 25.

(12) Leclere, Q.; Dinsenmeyer, A.; Antoni, J.; Julliard, E.; Pintado-Peño, A. Thresholded Multiple Coherence as a tool for source separation and denoising: Theory and aeroacoustic applications. *Applied Acoustics* **2021**, *178*, 108021.

(13) Allemang, R. J.; Patwardhan, R. S.; Kolluri, M. M.; Phillips, A. W. Frequency response function estimation techniques and the corresponding coherence functions: A review and update. *Mechanical Systems and Signal Processing* **2022**, *162*, 108100. DOI: https://doi.org/10.1016/j.ymssp.2021.108100.

(14) Eisenberg, R. S. Structural complexity, circuit models, and ion accumulation. *Fed Proc* **1980**, *39* (5), 1540-1543.

(15) Mathias, R. T. Analysis of Membrane Properties Using Extrinsic Noise. In *Membranes, Channels, and Noise*, Eisenberg, R., Frank, M., Stevens, C. Eds.; Springer US, 1984; pp 49-116.

(16) Eisenberg, R. Meeting Doug Henderson. *Journal of Molecular Liquids* **2022**, *361*, 119574. DOI: https://doi.org/10.1016/j.molliq.2022.119574. Eisenberg, B. Asking biological questions of physical systems: The device approach to emergent properties. *Journal of Molecular Liquids* **2018**, *270*, 212-217. Preprint available on arXiv as https://arxiv.org/abs/1801.05452.

(17) Eisenberg, B. Electrical Structure of Biological Cells and Tissues: impedance spectroscopy, stereology, and singular perturbation theory. In *Impedance Spectroscopy: Theory, Experiment, and Applications. Third Edition*, Third ed.; Barsoukov, E., Macdonald, J. R. Eds.; Wiley-Interscience, 2018; pp 472-478 Available on arXiv at https://arxiv.org/abs/1511.01339 as arXiv:01511.01339.

(18) Eisenberg, B. Setting Boundaries for Statistical Mechanics, Version 2: MDPI Molecules. *Molecules* **2022**, *27*, 8017. DOI: https://doi.org/10.3390/molecules27228017.

(19) Eisenberg, R. Structural Analysis of Fluid Flow in Complex Biological Systems. *Modeling and Artificial Intelligence in Ophthalmology.* **2023 4:**1-10 doi: 10.35119/maio.v4i1.126

*Preprints 2022, 2022050365 (doi: 10.20944/preprints202205.0365.v1).* **2022**. DOI: doi: 10.20944/preprints202205.0365.v1).

(20) Phillips, J. C.; Braun, R.; Wang, W.; Gumbart, J.; Tajkhorshid, E.; Villa, E.; Chipot, C.; Skeel, R. D.; Kale, L.; Schulten, K. Scalable molecular dynamics with NAMD. *J Comput Chem* **2005**, *26* (16), 1781-1802. DOI: 10.1002/jcc.20289. Leimkuhler, B. J.; Reich, S.; Skeel, R. D. Integration methods for molecular dynamics. In *Mathematical Approaches to biomolecular structure and dynamics*, Mesirov, J. P., Schulten, K., Sumners, D. W. Eds.; Springer, 1996; pp 161-185.

(21) Brehm, M.; Kirchner, B. TRAVIS - A Free Analyzer and Visualizer for Monte Carlo and Molecular Dynamics Trajectories. *Journal of chemical information and modeling* **2011**, *51* (8), 2007-2023. DOI: 10.1021/ci200217w.

(22) Brehm, M.; Thomas, M.; Gehrke, S.; Kirchner, B. TRAVIS—A free analyzer for trajectories from molecular simulation. *The Journal of chemical physics* **2020**, *152* (16), 164105.

(23) Pauling, L.; Corey, R. B.; Branson, H. R. The structure of proteins; two hydrogen-bonded helical configurations of the polypeptide chain. *Proceedings of the National Academy of Sciences of the United States of America* **1951**, *37* (4), 205-211. DOI: 10.1073/pnas.37.4.205 PubMed. Eisenberg, D. The discovery of the α-helix and β-sheet, the principal structural features of proteins. *Proceedings of the National Academy of Sciences* **2003**, *100* (20), 11207-11210.

(24) Hendrickson, W. A.; Teeter, M. M. Structure of the hydrophobic protein crambin determined directly from the anomalous scattering of sulphur. *Nature* **1981**, *290* (5802), 107-113. DOI: 10.1038/290107a0.

(25) Liu, Y.; Beveridge, D. L. Exploratory studies of ab initio protein structure prediction: multiple copy simulated annealing, AMBER energy functions, and a generalized born/solvent accessibility solvation model. *Proteins* **2002**, *46* (1), 128-146. DOI: 10.1002/prot.10020 [pii]. Lange, A.; Giller, K.; Pongs, O.; Becker, S.; Baldus, M. Two-dimensional solid-state NMR applied to a chimeric potassium channel. *J Recept Signal Transduct Res* **2006**, *26* (5-6), 379-393. DOI: HP4P411022G408LH [pii]

10.1080/10799890600932188. Orellana, L. Large-Scale Conformational Changes and Protein Function: Breaking the in silico Barrier. *Frontiers in Molecular Biosciences* **2019**, *6*. DOI: 10.3389/fmolb.2019.00117.

(26) Bang, D.; Tereshko, V.; Kossiakoff, A. A.; Kent, S. B. H. Role of a salt bridge in the model protein crambin explored by chemical protein synthesis: X-ray structure of a unique protein analogue, [V15A]crambin-α-carboxamide. *Molecular BioSystems* **2009**, *5* (7), 750. DOI: 10.1039/b903610e.

(27) Schmidt, A.; Teeter, M.; Weckert, E.; Lamzin, V. S. Crystal structure of small protein crambin at 0.48 Å resolution. *Acta Crystallogr Sect F Struct Biol Cryst Commun* **2011**, *67* (Pt 4), 424-428. DOI: 10.1107/S1744309110052607 PubMed. Chen, J. C. H.; Fisher, Z.; Kovalevsky, A. Y.; Mustyakimov, M.; Hanson, B. L.; Zhurov, V. V.; Langan, P. Room-temperature ultrahigh-resolution time-of-flight neutron and X-ray diffraction studies of H/D-exchanged crambin. *Acta Crystallogr Sect F Struct Biol Cryst Commun* **2012**, *68* (Pt 2), 119-123. DOI: 10.1107/S1744309111051499 PubMed.

(28) Kittel, C. *Solid-State Physics, Eighth Edition*; Wiley, 2004. Ferry, D. K. *An Introduction to Quantum Transport in Semiconductors*; Jenny Stanford Publishing, 2017. Shockley, W. *Electrons and Holes in Semiconductors to applications in transistor electronics*; van Nostrand, 1950. Sze, S. M. *Physics of Semiconductor Devices*; John Wiley & Sons, 1981.

(29) Cardona, M. Albert Einstein as the father of solid state physics. *arXiv preprint physics/0508237* **2005**.

(30) Hollander, M.; Wolfe, D. A.; Chicken, E. *Nonparametric Statistical Methods*; Wiley, 2013. Peter, H.; Nicholas, I. F.; Branka, H. On the Nonparametric Estimation of Covariance Functions. *The Annals of Statistics* **1994**, *22* (4), 2115-2134. DOI: 10.1214/aos/1176325774.

(31) Eisenberg, R. S. Computing the field in proteins and channels. *Journal of Membrane Biology* **1996**, *150*, 1–25. Preprint available on physics arXiv as document 1009.2857. Eisenberg, R. S. Atomic Biology, Electrostatics and Ionic Channels. In *New Developments and Theoretical Studies of Proteins*, Elber, R. Ed.; Vol. 7; World Scientific, 1996; pp 269-357. Published in the Physics ArXiv as arXiv:0807.0715.

(32) Maki, B. E. Interpretation of the coherence function when using pseudorandom inputs to identify nonlinear systems. *IEEE transactions on biomedical engineering* **1986**, (8), 775-779.

(33) Brigham, E. O. *Fast Fourier Transform and its Applications*; Prentice Hall, 1988. Briggs, W. L.; Henson, V. E. *The DFT: An Owner's Manual for the Discrete Fourier Transform*; SIAM, 1995.

(34) Li, Y. F.; Chen, K. F. Eliminating the picket fence effect of the fast Fourier transform. *Comput Phys Commun* **2008**, *178* (7), 486-491. DOI: https://doi.org/10.1016/j.cpc.2007.11.005. Chen, K. F.; Jiang, J. T.; Crowsen, S. Against the long-range spectral leakage of the cosine window family. *Comput Phys Commun* **2009**, *180* (6), 904-911. Chen, K. F.; Li, Y. F. Combining the Hanning windowed interpolated FFT in both directions. *Comput Phys Commun* **2008**, *178* (12), 924-928.

(35) Pauling, L. The nature of the chemical bond. Application of results obtained from the quantum mechanics and from a theory of paramagnetic susceptibility to the structure of molecules. *Journal of the American Chemical Society* **1931**, *53* (4), 1367-1400. Pauling, L. *Nature of the Chemical Bond*; Cornell University Press, 1939.

(36) Borwein, D.; Borwein, J. M.; Taylor, K. F. Convergence of lattice sums and Madelung's constant. *Journal of mathematical physics* **1985**, *26* (11), 2999-3009.

(37) Glasser, L. Solid-State Energetics and Electrostatics: Madelung Constants and Madelung Energies. *Inorganic Chemistry* **2012**, *51* (4), 2420-2424. DOI: 10.1021/ic2023852.

(38) Eisenberg, R. A Necessary Addition to Kirchhoff's Current Law of Circuits, Version 2. *Engineering Archive EngArXiv* **2022**, *https://doi.org/10.31224/2234*. DOI: https://doi.org/10.31224/2234. Eisenberg, R. S. Electrodynamics Correlates Knock-on and Knock-off: Current is Spatially Uniform in Ion Channels. *Preprint on arXiv at https://arxiv.org/abs/2002.09012* **2020**.

(39) McClellan, J. H.; Schafer, R. W.; Yoder, M. A. *Dsp first*; Pearson Education, 2017.

(40) Karlin, S.; Taylor, H. M., New York. *A First Course in Stochastic Processes*; Academic Press, 1975. Karlin, S.; Taylor, H. M. *A Second Course in Stochastic Processes*; Academic Press, 1981. Schuss, Z. *Theory And Applications Of Stochastic Processes: An Analytical Approach*; Springer, 2009.

(41) Muldowney, P. *A Modern Theory of Random Variation*; Wiley, 2012.

(42) Muldowney, P. *Gauge Integral Structures for Stochastic Calculus and Quantum Electrodynamics*; Wiley, 2021.

(43) Eisenberg Robert, S. Impedance Measurement of the Electrical Structure of Skeletal Muscle. In *Comprehensive Physiology, Republished by the American Physiological Society, as Volume 1, Supplement 27 of Handbook of Physiology, part of Comprehensive Physiology, SSN: 20404603 Online ISBN: 9780470650714*, Vol. 1, Supplement 27, Handbook of Physiology; American Physiological Society and Wiley OnLine, 2011.

(44) Eisenberg, R. S.; Mathias, R. T. Structural analysis of electrical properties of cells and tissues. *CRC Crit Rev Bioeng* **1980**, *4* (3), 203-232.

(45) Mandel, L.; Wolf, E. *Optical coherence and quantum optics*; Cambridge university press, 1995. Wolf, E. Unified theory of coherence and polarization of random electromagnetic beams. *Physics letters A* **2003**, *312* (5-6), 263-267. Wolf, E. Introduction to the Theory of Coherence and Polarization of Light. Cambridge University Press, 2007; p 235.

(46) Lanczos, C. *Linear Differential Operators*; Dover Publications, 1997. Lanczos, C. *Applied analysis*; Courier Corporation, 1988.

(47) Klee, H.; Allen, R. *Simulation of Dynamic Systems with MATLAB and Simulink*; CRC Press, 2018. Marple, S. L. *Digital Spectral Analysis MATLAB Software User Guide*; Dover Publications, 2019. ElAli, T. S. *Continuous Signals and Systems with MATLAB*; Crc Press, 2020.

(48) MATLAB. 9.3 (R2017b); The MathWorks Inc.: Natick, assachusetts, 2017.

(49) Otnes, R. K.; Enochson, L. *Digital Time Series Analysis*; John Wiley, 1972.

(50) Tervo, J.; Setälä, T.; Friberg, A. T. Theory of partially coherent electromagnetic fields in the space–frequency domain. *J. Opt. Soc. Am. A* **2004**, *21* (11), 2205-2215. DOI: 10.1364/JOSAA.21.002205. Tervo, J.; Setälä, T.; Friberg, A. T. Degree of coherence for electromagnetic fields. *Opt. Express* **2003**, *11* (10), 1137-1143. DOI: 10.1364/OE.11.001137. Price, S.; Bernhard, R. Virtual coherence: A digital signal processing technique for incoherent source identification. In *Proceedings of IMAC*, 1986; Vol. 4, pp 3-6. Suki, B.; Lutchen, K. Pseudorandom signals to estimate apparent transfer and coherence functions of nonlinear systems: applications to respiratory mechanics. *IEEE transactions on biomedical engineering* **1992**, *39* (11), 1142-1151.

(51) Banwell, C. N.; McCash, E. M. *Fundamentals of molecular spectroscopy*; McGraw-Hill New York, 1994. Barsoukov, E.; Macdonald, J. R. *Impedance spectroscopy: theory, experiment, and applications*; John Wiley & Sons, 2018. Demchenko, A. P. *Ultraviolet spectroscopy of proteins*; Springer Science & Business Media, 2013. Kremer, F.; Schönhals, A. *Broadband Dielectric Spectroscopy*; Springer, 2003. Rao, K. N. *Molecular spectroscopy: modern research*; Elsevier, 2012. Stuart, B. *Infrared spectroscopy*; Wiley Online Library, 2005.

(52) Parsegian, V. A. *Van der Waals Forces: A Handbook for Biologists, Chemists, Engineers, and Physicists*; Cambridge University Press, 2006.

(53) Oncley, J.; Ferry, J.; Shack, J. The measurement of dielectric properties of protein solutions; a discussion of methods and interpretation. In *Cold Spring Harbor Symposia on Quantitative Biology*, 1938; Cold Spring Harbor Laboratory Press: Vol. 6, pp 21-23. Oncley, J. L. Dielectric behavior and atomic structure of serum albumin. *Biophys Chem* **2003**, *100* (1-3), 151-158. DOI: S0301462202002764 [pii].

(54) Sun, H. H. *Synthesis of RC Networks*; Hayden Book Publishers, 1967. Tuttle, D. F. *Network synthesis*; Wiley, 1958. Weinberg, L. *Network analysis and synthesis*; Krieger Pub. Co., 1975. Zemanian, A. H. *Infinite electrical networks*; Cambridge University Press, 1991. Swanson, D. C. *Signal processing for intelligent sensor systems with MATLAB*; CRC Press, 2011. Hughes, T. H.; Morelli, A.; Smith, M. C. Electrical network synthesis: A survey of recent work. In *Emerging Applications of Control and Systems Theory*, Springer, 2018; pp 281-293.

(55) Eisenberg, R. Electrical Structure of Biological Cells and Tissues. In *Impedance spectroscopy: theory, experiment, and applications*, Barsoukov, E., Macdonald, J. R. Eds.; John Wiley, 2018.

(56) Tin-Lam, T.; Tuan-Seng, C. On Itô-Kurzweil-Henstock integral and integration-by-part formula. *Czechoslovak Mathematical Journal* **2005**, *55* (3), 653-663. Kurtz, D. S.; Swartz, C. W. *Theories of integration: the integrals of Riemann, Lebesgue, Henstock-Kurzweil, and Mcshane*; World Scientific Publishing Company, 2004.

(57) Toh, T.-L.; Chew, T.-S. The Kurzweil-Henstock theory of stochastic integration. *Czechoslovak Mathematical Journal* **2012**, *62* (3), 829-848. DOI: 10.1007/s10587-012-0048-z.

(58) Lab, O. Coherence and Correlation 4.3.9. *https://www.originlab.com/doc/Tutorials/Coherence-and-Correlation* **2022**.

(59) Simulink. *Simulink Reference Guide*; The MathWorks, 2022.

(60) Eisenberg, R. S.; Mathias, R. T.; Rae, J. S. Measurement, modeling, and analysis of the linear electrical properties of cells. *Ann N Y Acad Sci* **1977**, *303*, 342-354.

(61) Amidror, I. *Mastering the discrete Fourier transform in one, two or several dimensions: pitfalls and artifacts*; Springer, 2013.

(62) Temes, G. C.; Barcilon, V.; Marshall, F. C. The optimization of bandlimited systems. *Proceedings of the IEEE* **1973**, *61*, 196-234.

(63) Taylor, J. *Classical Mechanics*; University Science Books, 2005.

(64) Søndergaard, C. R.; Olsson, M. H.; Rostkowski, M.; Jensen, J. H. Improved treatment of ligands and coupling effects in empirical calculation and rationalization of p K a values. *Journal of chemical theory and computation* **2011**, *7* (7), 2284-2295.

(65) Maier, J. A.; Martinez, C.; Kasavajhala, K.; Wickstrom, L.; Hauser, K. E.; Simmerling, C. ff14SB: improving the accuracy of protein side chain and backbone parameters from ff99SB. *Journal of chemical theory and computation* **2015**, *11* (8), 3696-3713.

(66) Izadi, S.; Onufriev, A. V. Accuracy limit of rigid 3-point water models. *The Journal of Chemical Physics* **2016**, *145* (7), 074501. DOI: 10.1063/1.4960175.

(67) Joung, I. S.; Cheatham, T. E. Determination of Alkali and Halide Monovalent Ion Parameters for Use in Explicitly Solvated Biomolecular Simulations. *The Journal of Physical Chemistry B* **2008**, *112* (30), 9020-9041, doi: 10.1021/jp8001614. DOI: 10.1021/jp8001614.

(68) Eastman, P.; Pande, V. OpenMM: A hardware-independent framework for molecular simulations. *Computing in science & engineering* **2010**, *12* (4), 34-39.

(69) Allen, M. P.; Tildesley, D. J. *Computer Simulation of Liquids*; Oxford, 1987.

(70) Bullerjahn, J. T.; Von Bülow, S.; Hummer, G. Optimal estimates of self-diffusion coefficients from molecular dynamics simulations. *The Journal of Chemical Physics* **2020**, *153* (2), 024116. DOI: 10.1063/5.0008312.

(71) von Bülow, S.; Bullerjahn, J. T.; Hummer, G. Systematic errors in diffusion coefficients from long-time molecular dynamics simulations at constant pressure. *The Journal of Chemical Physics* **2020**, *153* (2), 021101.

(72) Humphrey, W.; Dalke, A.; Schulten, K. VMD -Visual Molecular Dynamics. *Journal of Molecular Graphics* **1996**, *14*, 33-38.

(73) Brooks, B.; Karplus, M. Harmonic dynamics of proteins: normal modes and fluctuations in bovine pancreatic trypsin inhibitor. *Proceedings of the National Academy of Sciences* **1983**, *80* (21), 6571-6575. DOI: 10.1073/pnas.80.21.6571.

(74) Bauer, J. A.; Pavlović, J.; Bauerová-Hlinková, V. Normal mode analysis as a routine part of a structural investigation. *Molecules* **2019**, *24* (18), 3293.

(75) Tirion, M. M. Large amplitude elastic motions in proteins from a single-parameter, atomic analysis. *Physical review letters* **1996**, *77* (9), 1905.

(76) Arunan, E.; Desiraju, G. R.; Klein, R. A.; Sadlej, J.; Scheiner, S.; Alkorta, I.; Clary, D. C.; Crabtree, R. H.; Dannenberg, J. J.; Hobza, P. Definition of the hydrogen bond (IUPAC Recommendations 2011). *Pure and applied chemistry* **2011**, *83* (8), 1637-1641.

(77) Spiridon, L.; Sulea, T. A.; Minh, D. D. L.; Petrescu, A. J. Robosample: A rigid-body molecular simulation program based on robot mechanics. *Biochim Biophys Acta Gen Subj* **2020**, *1864* (8), 129616. DOI: 10.1016/j.bbagen.2020.129616 From NLM Medline.

(78) Liwo, A. Coarse graining: a tool for large-scale simulations or more? *Physica Scripta* **2013**, *87* (5), 058502.

(79) Sikorska, C.; Liwo, A. Origin of Correlations between Local Conformational States of Consecutive Amino Acid Residues and Their Role in Shaping Protein Structures and in Allostery. *The Journal of Physical Chemistry B* **2022**, *126* (46), 9493-9505. DOI: 10.1021/acs.jpcb.2c04610.

# Supplemental Information

## Error Estimates

**Stochastic errors** can be evaluated from the estimation procedures described in some detail in the text. These are nearly summarized in the Table below, reworked from Bendat and Piersol [1].

**Random Error Formulas**

Table S1

| Name | Function (symbol) | Random error |
|---|---|---|
| Coherence | $\hat{\gamma}^2_{xy}(f)$ | $\dfrac{\sqrt{2}\left[1 - \hat{\gamma}^2_{xy}(f)\right]}{\hat{\gamma}_{xy}(f)\sqrt{n_d}}$ |
| Power | $\hat{G}_{xx}(f) = \hat{\gamma}^2_{xy}(f)\hat{G}_{yy}(f)$ | $\dfrac{\left[2 - \hat{\gamma}^2_{xy}(f)\right]^{\frac{1}{2}}}{\left|\hat{\gamma}_{xy}(f)\right|\sqrt{n_d}}$ |
| Magnitude of Frequency Function | $\left|\hat{H}_{xy}(f)\right|$ | $\dfrac{\left[1 - \hat{\gamma}^2_{xy}(f)\right]^{\frac{1}{2}}}{\left|\hat{\gamma}_{xy}(f)\right|\sqrt{2n_d}}$ |
| Phase of Frequency Function | $\hat{\varphi}_{xy}(f)$ | $SD\left[\hat{\varphi}_{xy}(f)\right] = \dfrac{\left[1 - \hat{\gamma}^2_{xy}(f)\right]^{\frac{1}{2}}}{\left|\hat{\gamma}_{xy}(f)\right|\sqrt{2n_d}}$ |

$n_d = number\ of\ data\ points$; $SD$ = standard deviation

Modified version of Bendat & Piersol[1,] Table 9.6, p. 312

**Systematic Error**

Most systematic errors of concern arise from details in the computations of molecular dynamics and estimation of the parameters of Table S1. These are discussed in the main text.

The comparison (see Figures 3-6 of text) of analytical expressions and stochastic estimation of the model system defined in Figure 2 suggest that systematic errors are not present in large amounts. The finding that non-interacting atoms have very low coherence (Figure 9) gives comfort that whatever errors remain do not create coherence out of nothing.

Further checks were made to show that displacement of the structure did not change estimates of coherence.

First (Figure S1) the coherence of the crambin protein was estimated after 100 uniformly sampled displacements. That is, we translated every atom in the protein by displacement uniformly sampled between $-200$ and $+200$ Angstroms.

The results shown are the coherences of a particular H-bond (PHE13N and ALA9O) in the main alpha helix but every atom examined showed similar results. We see slight deviations in coherence for mostly higher frequencies, where the mean coherence itself is less reliable. Coherence at lower frequencies are $> 0.9$ despite displacements.

## References in Supplemental Material

[1]  J. S. Bendat and A. G. Piersol, *Random data: analysis and measurement procedures Fourth Edition*. (John Wiley & Sons, 2010).

# Figure S1

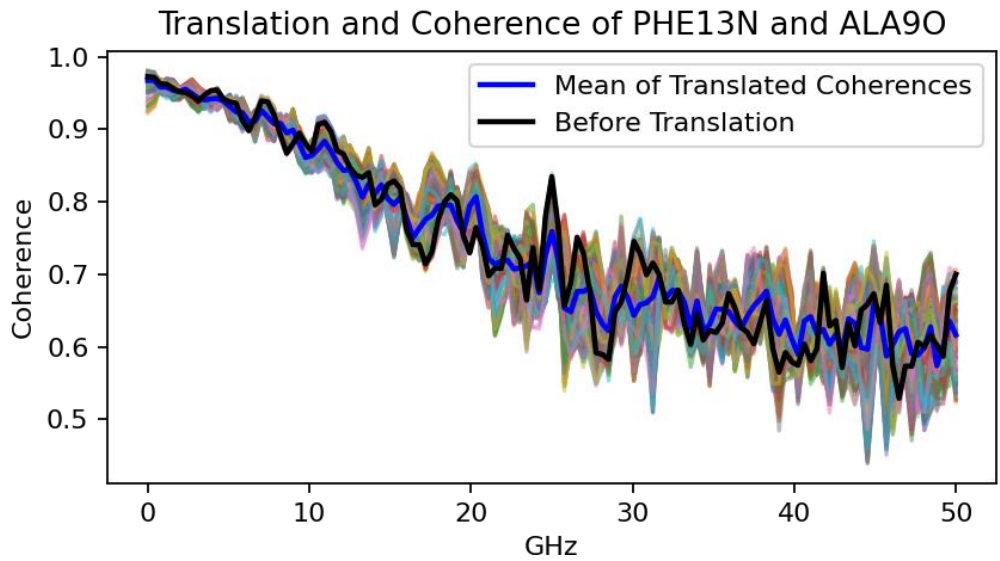

Figure S1: The black line is the coherence calculated from the native trajectory and the average of all the translated coherences is shown in blue. We see there is not much discrepancy between the averaged (blue) coherence and the coherence calculated from the native trajectory, especially in the low frequency range.

# Figure S2

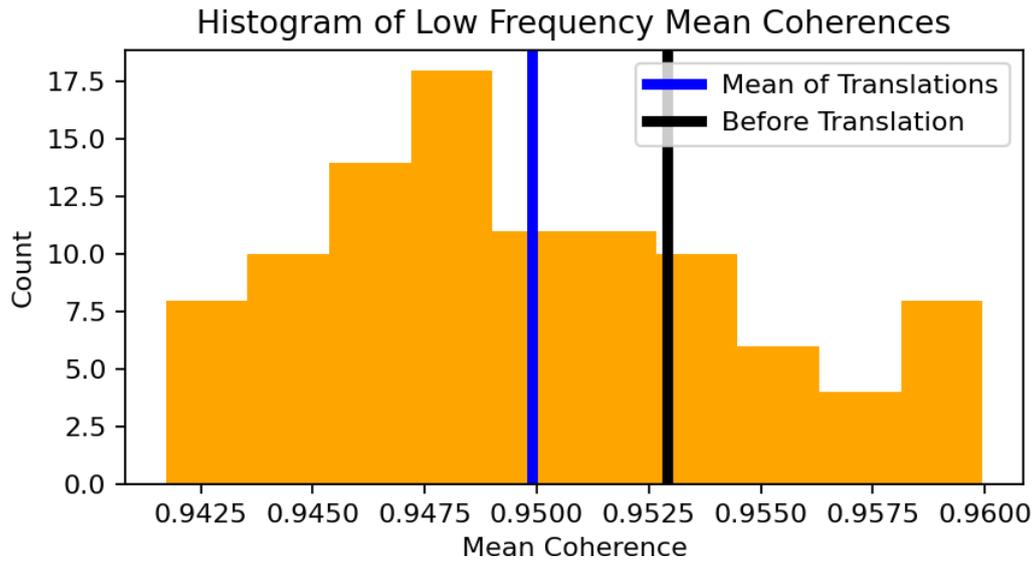

Figure S2: Histogram of the averaged coherence between 0.39 GHz and 5.09 GHz (the main range of interest) of all the translated trajectories. Each trajectory was separately translated by different amounts, as specified above. The standard deviation of $4.69 \cdot 10^{-3}$ in the mean coherence suggests that our coherence estimation is not sensitive to translation.

# Figure S3

# Rotating the protein does not change coherence.

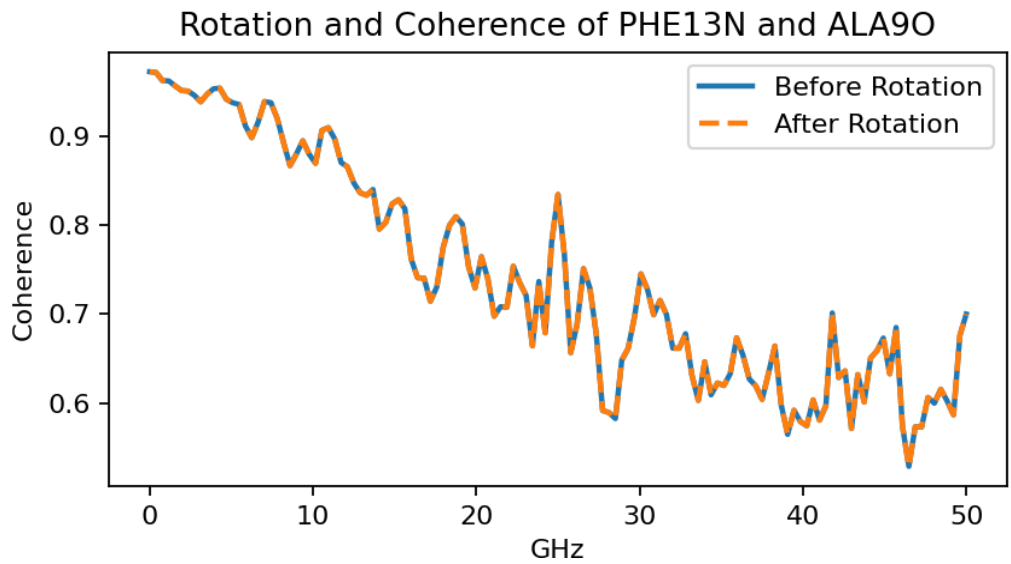